\documentclass[unnumsec,webpdf,contemporary,large]{oup-authoring-template}%

\graphicspath{{Fig/}}

\theoremstyle{thmstyleone}%
\theoremstyle{thmstyletwo}%
\theoremstyle{thmstylethree}%

\begin{document}

\journaltitle{arXiv}
\DOI{Preprint}
\copyrightyear{2021}
\pubyear{2021}
\access{OmiEmbed}
\appnotes{Problem Solving Protocol}

\firstpage{1}


\title[OmiEmbed]{OmiEmbed: a unified multi-task deep learning framework for multi-omics data}

\author[1,$\ast$]{Xiaoyu Zhang}
\author[1]{Yuting Xing}
\author[1]{Kai Sun}
\author[1,2,$\ast$]{Yike Guo}

\authormark{Zhang et al.}

\address[1]{\orgdiv{Data Science Institute}, \orgname{Imperial College London}, \orgaddress{\postcode{SW7 2AZ}, \state{London}, \country{UK}}}
\address[2]{\orgdiv{Department of Computer Science}, \orgname{Hong Kong Baptist University}, \orgaddress{\state{Hong Kong}, \country{China}}}

\corresp[$\ast$]{Corresponding author. \href{x.zhang18@imperial.ac.uk}{x.zhang18@imperial.ac.uk}}

\received{Date}{0}{Year}
\revised{Date}{0}{Year}
\accepted{Date}{0}{Year}



\abstract{High-dimensional omics data contains intrinsic biomedical information that is crucial for personalised medicine. Nevertheless, it is challenging to capture them from the genome-wide data due to the large number of molecular features and small number of available samples, which is also called `the curse of dimensionality' in machine learning. To tackle this problem and pave the way for machine learning aided precision medicine, we proposed a unified multi-task deep learning framework named OmiEmbed to capture biomedical information from high-dimensional omics data with the deep embedding and downstream task modules. The deep embedding module learnt an omics embedding that mapped multiple omics data types into a latent space with lower dimensionality. Based on the new representation of multi-omics data, different downstream task modules were trained simultaneously and efficiently with the multi-task strategy to predict the comprehensive phenotype profile of each sample. OmiEmbed support multiple tasks for omics data including dimensionality reduction, tumour type classification, multi-omics integration, demographic and clinical feature reconstruction, and survival prediction. The framework outperformed other methods on all three types of downstream tasks and achieved better performance with the multi-task strategy comparing to training them individually. OmiEmbed is a powerful and unified framework that can be widely adapted to various application of high-dimensional omics data and has a great potential to facilitate more accurate and personalised clinical decision making.}


\maketitle

\section{Introduction}
With the increasingly massive amount of omics data generated from emerging high-throughput technologies, the large-scale, cost-efficient and comprehensive analysis of biological molecules becomes an everyday methodology for biomedical researchers \citep{Hasin2017MultiomicsAT, Berger2013ComputationalSF}. The analysis and assessment of different types of omics data facilitate the integration of molecular features during the standard diagnostic procedure. For instance, in the 2016 World Health Organization (WHO) classification of central nervous system (CNS) tumours \citep{louis20162016} an integrative method combining both histopathology and molecular information was recommended for the identification of multiple tumour entities. Nevertheless, most of these molecular features designed to aid diagnosis are manually selected biomarkers referring to specific genetic alterations, which neglects the genome-wide patterns correlated with disease prognosis and other phenotypic outcomes. In recent years, instead of focusing on the effect of specific molecular features, many researchers began to delve into the overall picture of genome-wide omics data and try to obtain deep understanding of diseases and uncover crucial diagnostic or prognostic information from it \citep{Chaudhary2017DeepLM, capper2018dna,Zhang2019IntegratedMA,Amodio2019ExploringSD}. 

It is challenging to analyse genome-wide high dimensional omics data because of the mismatch between the number of molecular features and the number of samples. The dimensionality of genome-wide omics data is fairly high. For example, a RNA-Seq gene expression profile is consisted of more than 60,000 identifiers, and a HumanMethylation450 (450K) DNA methylation profile has more than 485,000 probes, while the number of available samples in an omics dataset is normally small due to the difficulty of patient recruitment and sample collection. This phenomenon is called ``the curse of dimensionality'' in machine learning which would cause massive overfitting of a model and make samples hard to cluster \citep{goodfellow2016deep}. To overcome this issue, the number of molecular features used in downstream tasks is required to reduce significantly. Two common approaches are 1) to manually select a subset of the molecular features related to the downstream task based on domain knowledge; 2) to apply traditional dimensionality reduction algorithms, e.g., principal component analysis (PCA).

Inspired by the significant success in fields like computer vision \citep{Voulodimos2018DeepLF} and natural language processing \citep{Young2018RecentTI}, deep learning approaches have been applied to analyse the complicated and nonlinear relationships between molecular features of high-dimensional omics data and detect genome-wide biological patterns from them \citep{Ding2018InterpretableDR, Lopez2018DeepGM, Eraslan2019SinglecellRD}. With the deep learning mechanism, molecular features can be automatically selected during the training process without manual selection. Multiple downstream tasks were conducted on different types of high-dimensional omics data, including dimensionality reduction \citep{Ding2018InterpretableDR, Way2018ExtractingAB}, disease type classification \citep{Zhang2019IntegratedMA, Ma2019AffinityNetSF}, survival prediction \citep{Chaudhary2017DeepLM, Cheerla2019DeepLW}. However, there is no unified deep learning method, to the best of our knowledge, that can simultaneously conduct all aforementioned downstream tasks together on any combination of omics types.

Here we proposed a unified multi-task deep learning framework for integrated multi-omics analysis named OmiEmbed. The OmiEmbed framework is comprised of two main components: deep embedding module and downstream task module. In the deep embedding module, high-dimensional multi-omics data was embedded into a low-dimensional latent space to tackle the challenge of `dimensionality curse'. The learnt novel representation of each sample was then fed to multiple downstream networks which were trained simultaneously with a joint loss function and the multi-task training strategy. Different downstream tasks that were already implemented in OmiEmbed include tumour type classification, demographic and clinical feature (e.g., age, gender, race, primary site and disease stage of sample) reconstruction and prognosis prediction (i.e., predicting the survival function of each input sample). The model was trained and evaluated on two publicly available omics datasets, the Genomic Data Commons (GDC) pan-cancer multi-omics dataset \citep{grossman2016toward} and the GSE109381 brain tumour methylation (BTM) dataset \citep{capper2018dna}. Our model achieved promising results for all aforementioned downstream tasks and outperformed other corresponding existing methods. With the multi-task training strategy OmiEmbed was able to infer all downstream tasks simultaneously and efficiently. Better results were achieved in the multi-task scenario comparing to training and inferring each downstream task separately.

\section{Related work}
The representation learning ability of DNNs (deep neural networks) has been widely verified by the significant breakthrough in computer vision and natural language processing. Inspired by this achievement, a number of deep learning approaches have been applied to high-dimensional omics data for different downstream tasks in recent years. 

The most common downstream task is classification. \cite{Danaee2017ADL} presented a cancer detection model that discriminated breast tumour samples from normal samples using gene expression data based on a stacked denoising autoencoder (SDAE). \cite{Lyu2018DeepLB} reshaped the high-dimensional RNA-Seq data into images and applied a convolutional neural network (CNN) for tumour type classification on the GDC dataset, which got the accuracy of 95.59\%. \cite{Rhee2018HybridAO} proposed a hybrid model that was comprised of a graph convolution neural network (GCN) and a relation network (RN) for breast tumour subtype classification using gene expression data and protein-protein interaction (PPI) networks. \cite{Jurmeister2019MachineLA} developed a multi-layer neural network to distinguish metastatic head and neck squamous cell carcinoma (HNSC) from primary squamous cell carcinoma of the lung (LUSC) with an accuracy of 96.4\% in the validation cohort. The AffinityNet \citep{Ma2019AffinityNetSF} was a data efficient deep learning model that comprised multiple stacked k-nearest neighbours (KNN) attention pooling layers for tumour type classification. OmiVAE \citep{Zhang2019IntegratedMA} was an end-to-end deep learning method designed for tumour type classification based on a deep generative model variational autoencoder (VAE) achieving an accuracy of 97.49\% among 33 tumour types and the normal control using gene expression and DNA methylation data from the GDC dataset.

Another typical task that has been tackled by deep learning approaches recently is the prediction of prognosis status from high-dimensional omics data. \cite{Chaudhary2017DeepLM} applied a vanilla autoencoder (AE) to reduce the dimensionality of multi-omics data which was comprised of gene expression, DNA methylation and miRNA expression profiles and used the learned representation to identify two different survival subgroups of liver tumours by Cox proportional hazard model (CoxPH), K-means clustering and support vector machine (SVM). In their experiment, a concordance index (C-index) of 0.68 was achieved on the liver tumour subjects from the GDC dataset. The deep learning model applied in this research was not an end-to-end model, and the embedding learned by the network was used separately outside the network for downstream tasks. \cite{Huang2019SALMONSA} implemented a deep learning network with the CoxPH model to predict prognosis for breast tumour using multi-omics data, cancer biomarkers and a gene co-expression network. Aforementioned research focused mostly on tumour samples of specific primary site and neglected the information cross different tumour types which had the potential to improve the performance of survival prediction for each tumour type. \cite{Cheerla2019DeepLW} constructed a multimodal deep learning network to predict survival of subjects for 20 different tumour types in the GDC dataset which achieving an overall C-index of 0.78 based on additional clinical information and histopathology whole slide images (WSIs) besides the multi-omics data.

There are also several attempts on applying deep learning methodology to multiple downstream tasks for high-dimensional omics data. \cite{Amodio2019ExploringSD} presented a deep neural network method named SAUCIE to explore single-cell gene expression data and perform multiple data analysis task including clustering, batch correction, imputation and denoising, and visualisation. However, the backbone of SAUCIE was basically an autoencoder used for embedding learning, and most of the downstream tasks were required to conduct outside the network separately, hence the network was not able to perform all of the tasks simultaneously with a single training process. Deepathology \citep{Azarkhalili2019DeePathologyDM} was another deep learning method for omics data analysis which adopted the idea of multi-task learning. This model encoded gene expression profile into a low-dimensional latent vector to predict the tumour type and primary site of the input sample, which obtained an accuracy of 98.1\% for primary site prediction and 95.2\% for tumour type classification. In spite of the good results on multiple classification tasks, Deepathology was not able to perform the more complicated survival prediction task and did not adopt any state-of-the-art deep multi-task learning optimisation mechanism.

\section{Materials and methods}

\subsection{Datasets}
Two publicly available datasets were used as examples to demonstrate the ability of OmiEmbed: the Genomic Data Commons (GDC) pan-cancer multi-omics dataset \citep{grossman2016toward} and the DNA methylation dataset of human central nervous system tumours (GSE109381) \citep{capper2018dna}.

The GDC pan-cancer dataset is one of the most comprehensive and widely used multi-omics datasets. It comprises high-dimensional omics data and corresponding phenotype data from two cancer genome programmes: The Cancer Genome Atlas (TCGA) \citep{weinstein2013cancer} and Therapeutically Applicable Research to Generate Effective Treatment (TARGET). The TARGET programme mainly focuses on pediatric cancers. Three types of omics data from the GDC dataset were used in our experiments, including RNA-Seq gene expression profiling, DNA methylation profiling and miRNA expression profiling. The dimensionalities of the three types of omics data are 60,483, 485,577 and 1,881 respectively. This dataset consists of 36 different types of tumour samples along with corresponding normal control samples, among which 33 tumour types are from TCGA and 3 tumour types are from TARGET. The detailed tumours type information is tabulated in Supplementary Table S1. A wide range of phenotype features are also available in the GDC dataset including demographics (e.g., age, gender, and race), clinical sample information (e.g., primary site and disease stage of the sample) and the survival information (recorded time of death or censoring).

The GSE109381 brain tumour methylation (BTM) dataset from the Gene Expression Omnibus (GEO) is one of the largest DNA methylation datasets specifically targeted brain tumours. We integrated both the reference set and validation set of this dataset and the whole dataset consists of 3,905 samples with almost all WHO-defined central nervous system (CNS) tumour entities \citep{louis20162016} and seven non-neoplastic control CNS regions. Genome-wide DNA methylation profile for each sample was generated using Infinium HumanMethylation450 BeadChip (450K) arrays, which is the same platform used for the GDC DNA methylation data. Each sample in this dataset has two types of diagnostic label, the histopathological class label defined by the latest 2016 WHO classification of CNS tumours \citep{louis20162016} and the methylation class label defined by the original paper of this dataset \citep{capper2018dna}. The detailed tumour type information of the two label systems was listed in Supplementary Table S8 and S9. Other phenotypic information is also available in this dataset including age, gender and the disease stage of each sample.

\subsection{Omics data process}
Raw data of the GSE109381 BTM dataset downloaded from GEO\footnote{\url{https://www.ncbi.nlm.nih.gov/geo/query/acc.cgi?acc=GSE109381}} were first processed by the Bioconductor R package minfi \citep{aryee2014minfi} to get the bate value of each CpG probe. Beta value is the ratio of methylated signal intensities and the overall signal intensities, which indicates the methylation level of a specific CpG site. The DNA methylation profile generated by the 450K array has 485,577 probes in total. Certain probes were removed during the feature filtering step according to the following criteria: probes targeting the Y chromosome (n = 416), probes containing the dbSNP132Common single-nucleotide polymorphism (SNP) (n = 7,998), probes not mapping to the human reference genome (hg19) uniquely (one mismatch allowed) (n = 3,965), probes not included in the latest Infinium MethylationEPIC BeadChip (EPIC) array (n = 32,260), the SNP assay probes (n = 65), the non-CpG loci probes (n = 3,091) and probes with missing values (N/A) in more than 10\% of samples (n = 2). We followed some of the criteria mentioned in the original paper of this dataset \citep{capper2018dna}. 46,746 probes were filtered out, which results in a final DNA methylation feature set of 438,831 CpG sites.

For the GDC pan-cancer dataset, the harmonised data of three omics types were downloaded from the UCSC Xena data portal\footnote{\url{https://xenabrowser.net/datapages/}} with the original data dimensionality. Each RNA-Seq gene expression profile is comprised of 60,483 identifiers referring to corresponding genes. The expression level is quantified by the Fragments Per Kilobase of transcript per Million mapped reads (FPKM) value, which has been $\log_{2}$-transformed. Feature filtering was applied to the gene expression data: targeting Y chromosome (n = 594) and zero expression in all samples (n = 1,904). In total, 2,440 genes were removed, leaving 58,043 molecular features for further analyses. As for miRNA expression profiles, the expression level of each miRNA stem-loop identifier was measured by the $\log_{2}$-transformed Reads Per Million mapped reads (RPM) value. All of the miRNA identifiers were kept in our experiments. For both the gene expression and miRNA expression profiles, the expression values were normalised to the range of 0 to 1 due to the input requirement of the OmiEmbed framework. The DNA methylation data in the GDC dataset were filtered based on the same criteria used for the BTM dataset. The remaining missing values in all datasets mentioned above were imputed by the mean of corresponding molecular features.

\subsection{Overall architecture}

\begin{figure}[!t]
    \centering
    \includegraphics[scale=0.79]{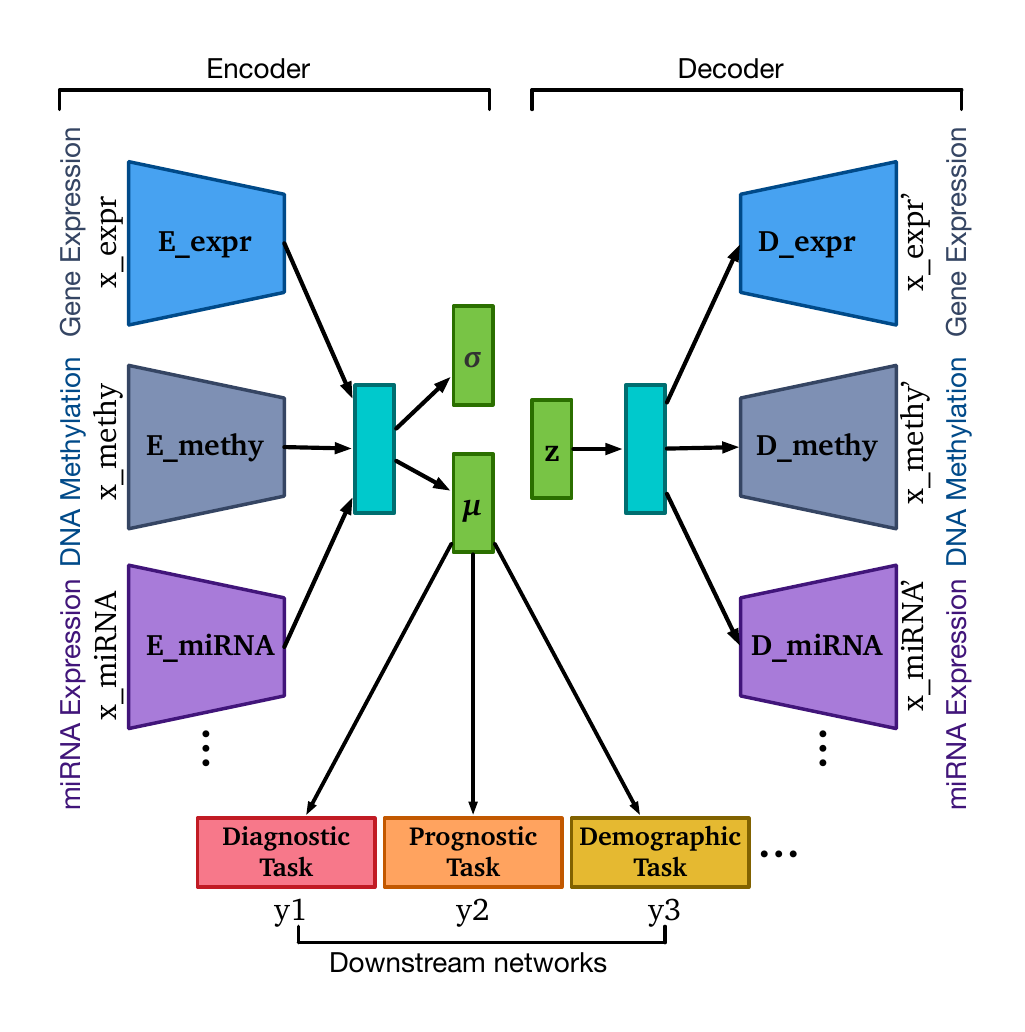}
    \caption{The overall architecture of OmiEmbed, which is comprised of two main components: the VAE deep embedding networks and the downstream task networks. The number of omics types and downstream tasks can be modified based on the user needs and requirements of the experiment. E\_expr, E\_methy and E\_miRNA represent encoders of gene expression, DNA methylation and miRNA expression respectively. Similarly, D\_expr, D\_methy and D\_miRNA represent decoders of gene expression, DNA methylation and miRNA expression.}
    \label{fig:network_structure}
\end{figure}

OmiEmbed is a unified end-to-end multi-view multi-task  deep learning framework designed for high-dimensional multi-omics data, with application to dimensionality reduction, tumour type classification, multi-omics integration, demographic and clinical feature reconstruction, and survival prediction. The overall architecture of OmiEmbed is comprised of a deep embedding module and one or multiple downstream task modules, which is illustrated in Fig. \ref{fig:network_structure}.

The role of the deep embedding module in OmiEmbed is to embed high-dimensional multi-omics profiles into a low-dimensional latent space for the downstream task modules. The backbone network we used in the deep embedding module is variational autoencoder (VAE) \citep{kingma2013auto}. VAE is a deep generative model which is also effective to capture the data manifold from high-dimensional data. We assume each sample $\mathbf{x}^{(i)} \in \mathbb{R}^d$  in the multi-omics dataset $\mathcal{D}$ can be represented by and generated from a latent vector $\mathbf{z}^{(i)} \in \mathbb{R}^p$, where $p \ll d$. In the generation process, each latent vector is first sampled from a prior distribution $p_\theta(\mathbf{z})$, and then the multi-omics data of each sample is generated from the conditional distribution $p_\theta(\mathbf{x|z})$, where $\theta$ is the set of learnable parameters of the decoder. In order to address the intractability of the true posterior $p_\theta(\mathbf{z|x})$, the variational distribution $q_{\phi}(\mathbf{z|x})$ is introduced to approximate $p_\theta(\mathbf{z|x})$, where $\phi$ is the set of learnable parameters of the encoder. As a result, the VAE network is optimised by maximizing the variational lower bound formularised as below:
\begin{equation}
    \mathbb{E}_{\mathbf{z} \sim q_{\phi}(\mathbf{z|x})} \log p_{\theta}(\mathbf{x|z}) - D_{\mathrm{KL}}\left(q_{\phi}(\mathbf{z|x}) \| p_\theta(\mathbf{z})\right)
\end{equation}
where $D_{\mathrm{KL}}$ is the Kullback-Leibler (KL) divergence between two probability distributions \citep{goodfellow2016deep}.

We applied the framework of VAE to our deep embedding module to obtain the low-dimensional latent vector that can represent the original high-dimensional omics data in the downstream task modules. For each type of omics data, the input profiles were first encoded into corresponding vectors with specific encoders. Those vectors of different omics types were then concatenated together in the subsequent hidden layer and encoded into one multi-omics vector. Based on the idea of VAE, the multi-omics vector was connected to two bottleneck layers in order to obtain the mean vector $\boldsymbol{\mu}$ and the standard deviation vector $\boldsymbol{\sigma}$. These two vectors defined the Gaussian distribution $\mathcal{N}\left(\boldsymbol{\mu}, \boldsymbol{\sigma}\right)$ of the latent variable $\mathbf{z}$ given the input sample $\mathbf{x}$, which is the variational distribution $q_{\phi}(\mathbf{z|x})$. Since sampling $\mathbf{z}$ from the learned distribution is not differentiable and suitable for backpropagation, the reparameterisation trick is applied as follows:
\begin{equation}
    \mathbf{z}=\boldsymbol{\mu}+\boldsymbol{\sigma} \boldsymbol{\epsilon}
\end{equation}
where $\boldsymbol{\epsilon}$ is a random variable sampled from the unit normal distribution $\mathcal{N}(\mathbf{0}, \mathbf{I})$. The latent variable $\mathbf{z}$ was then fed to the decoders with a symmetrical network structure to get the reconstructed multi-omics data $\mathbf{x}^\prime$.

We provided two types of detailed network structure for the encoders and decoders in the deep embedding module, the one-dimensional convolutional neural network (CNN) and the fully-connected neural network (FC). The network structures of the two types of deep embedding modules were illustrated in Supplementary Figure S1 and S2. Both network types shared the same architecture, and other state-of-the-art or customised embedding networks can be easily added to the OmiEmbed framework with minimal modification using our open-source repository in Github\footnote{\url{https://github.com/zhangxiaoyu11/OmiEmbed/}}. With the deep embedding module we can obtain the low-dimensional representation of the input omics data. This new representation can directly replace the original omics data as the input of any downstream task. For instance, when the latent dimension is set to 2 or 3 the new representation can be used for visualisation purpose. Nevertheless, we can also attach one or multiple downstream task networks to the bottleneck layer of the deep embedding module to get an end-to-end multi-task model, which is able to guide the embedding step with objectives and share information among different tasks.

Three main types of end-to-end downstream tasks were provided in the OmiEmbed framework: classification task, regression task and survival prediction task. Each downstream task fell into one of these categories can be trained individually or together with other downstream tasks using the coordinated multi-task strategy discussed in later sections. A multi-layer fully-connected network was applied to the classification-type downstream task, including diagnostic tasks such as tumour type classification, primary site prediction and disease stage (i.e., normal control, primary tumour, recurrent tumour or metastatic tumour) prediction and demographic tasks, e.g., the prediction of gender and race. The output dimension of the classification downstream network was set to the number of classes. For the regression task, a similar network was attached to the deep embedding module, but only one neuron was kept in the output layer to predict the target scalar value (e.g., age of the subject). The survival prediction downstream network is more complicated and addressed separately in a subsequent section. The downstream networks add further regularisation to the low dimensional latent representation and urge the deep embedding module to learn the representations related to certain downstream tasks. With the downstream modules, a single well-trained multi-task OmiEmbed network is able to reconstruct a comprehensive phenotype profile including diagnostic, prognostic and demographic information from omics data.

\subsection{Training strategy}
\label{sec:train_strategy}
The same as the overall structure, the joint loss function is also comprised of two main components: the loss of the deep embedding and the loss of the downstream tasks. 

We denote each type of input omics profile as $\mathbf{x}_{j}$ and the corresponding reconstructed profile as $\mathbf{x}_{j}^{\prime}$, where $j$ is the omics type index and there are $M$ omics types in total. The deep embedding loss can be then defined as follows:
\begin{equation}
    \mathcal{L}_{embed}=\frac{1}{M} \sum_{j=1}^{M} BCE \left(\mathbf{x}_{j}, \mathbf{x}_{j}^{\prime} \right) + D_{\mathrm{KL}}(\mathcal{N}(\boldsymbol{\mu}, \boldsymbol{\sigma}) \| \mathcal{N}(\mathbf{0}, \mathbf{I}))
\end{equation}
where $BCE$ is the binary cross-entropy to measure the distance between input data and reconstructed data, and the second term is the KL divergence between the learned distribution and a unit Gaussian distribution.

In the downstream modules, each downstream task has its specific loss function $\mathcal{L}_{down_k}$ and a corresponding weight $w_k$. For the classification type task, the loss function can be defined as:
\begin{equation}
    \mathcal{L}_{classification} = CE(y,y^{\prime})
\end{equation}
where $y$ is the ground truth, $y^{\prime}$ is the predicted label and $CE$ is the cross-entropy loss. Similar to the classification loss, the loss function of regression type task is
\begin{equation}
    \mathcal{L}_{regression} = MSE(y,y^{\prime})
\end{equation}
where $MSE$ is the mean squared error between the real value and the predicted value. The loss function of the survival prediction task is discussed separately in the next section. The overall loss function of the downstream modules is the weighted sum of all downstream losses, i.e.,
\begin{equation}
    \mathcal{L}_{down}=\frac{1}{K} \sum_{k=1}^{K} w_k \mathcal{L}_{down_k}
    \label{eq:down_loss}
\end{equation}
where $K$ is the number of downstream tasks, $\mathcal{L}_{down_k}$ is the loss for a certain task and $w_k$ is the corresponding weight. $w_k$ can be manually set as hyperparameters or used as learnable parameters during the training process. In conclusion, the joint loss function of the end-to-end OmiEmbed network is
\begin{equation}
    \mathcal{L}_{total} = \lambda \mathcal{L}_{embed} + \mathcal{L}_{down}
    \label{eq:total_loss}
\end{equation}
and depends on $\lambda$ that balance the two terms in the overall loss function.

Based on the aforementioned loss functions, three training phases were designed in OmiEmbed. Phase 1 was the unsupervised phase that only focused on the deep embedding module. In this training phase, only the deep embedding loss was backpropagated and only the parameters in the deep embedding network were updated based on the gradients. No label was required in the first training phase and this phase can be used separately as a dimensionality reduction or visualisation method. In Phase 2, the pre-trained embedding network was fixed whilst the downstream networks were being trained. The joint downstream loss was backpropagated and only the downstream networks were updated during this phase. After the embedding network and the downstream networks were pre-trained separately, the overall loss function defined in equation (\ref{eq:total_loss}) was calculated and backpropagated during Phase 3. In this final training phase the whole OmiEmbed network including the deep embedding module and the downstream modules was fine-tuned to obtain better performance.

\subsection{Survival prediction}
\label{sec:survival_method}
Survival prediction is the most complicated downstream task implemented in OmiEmbed. The objective of this task is to obtain individual survival function and hazard function for each subject based on the high-dimensional omics data. The survival function can be denoted by 
\begin{equation}
    S(t)=P[T>t]
\end{equation}
where $T$ is time elapsed between the sample collection time and the time of event occurring. The survival function signifies the probability that the failure event, i.e., death, has not occurred by time $t$. The hazard function can be defined as:
\begin{equation}
    h(t)=\lim _{dt \rightarrow 0} \frac{\operatorname{P}[t \leq T<t+dt \mid T \geq t]}{dt}
\end{equation}
which represents the instantaneous rate of occurrence for the failure event. A large hazard value manifests a great risk of death at specific time $t$. However, the original form of hazard function is infrequently used, and the risk score of each sample $\mathbf{x}$ is more commonly applied by subdividing the time axis into $m$ time intervals, such that:
\begin{equation}
    r(\mathbf{x})=\sum_{i=1}^{m} h\left(t_{i}, \mathbf{x}\right).
\end{equation}

In order to train a survival prediction downstream network, besides the omics data $\mathbf{x}$, two elements of the survival labels are required: the event time $T$ and the event indicator $E$. The indicator was set to 1 when the failure event was observed during the study and 0 when the event was not observed, which is termed censoring. In the case of censoring, the event time $T$ is the time elapsed between the time when the sample was collected and the time of the last contact with the subject. Both $T$ and $E$ are available in the GDC dataset.

The multi-task logistic regression (MTLR) model \citep{yu2011learning} was applied and adapted to the OmiEmbed framework for the survival prediction downstream task. In the first step, the time axis was divided into $m$ time intervals $\left\{l_i\right\}_{i=1}^{m}$. Each time interval was defined as $l_i=[t_{i-1}, t_i)$ where $t_0$ = 0 and $t_m \geq max(T)$. The number of time intervals $m$ is a hyperparameter. A larger $m$ results in more fine-grained output but requires more computation resources. We applied the multi-layer fully-connected network as the backbone of our survival prediction network and the dimension of the output layer is the number of time intervals. As a result, the output of our survival prediction network is an $m$-dimensional vector $\mathbf{y}^{\prime}=\left[y^{\prime}_{1}, y^{\prime}_{2}, \ldots, y^{\prime}_{m}\right]$. Similarly, the survival label of each subject was also encoded into an $m$-dimensional vector $\mathbf{y}=\left[y_{1}, y_{2}, \ldots, y_{m}\right]$, where $y_i$ signifies the survival status of this subject at the time point $t_i$. The likelihood of observing $\mathbf{y}$ on the condition of sample $\mathbf{x}$ with the network parameters $\theta$ can be defined as follows:
\begin{equation}
    P_{\theta}\left(\mathbf{y} \mid \mathbf{x}\right)=\frac{\exp \left(\sum_{i=1}^{m} y_{i} y^{\prime}_{i}\right)}{\sum_{j=0}^{m} \exp(\sum_{i=j+1}^{m} y^{\prime}_{i})}.
\end{equation}

The objective of this survival network is to find a set of parameters $\theta$ that maximises the log-likelihood, hence the loss function of the survival prediction downstream task is defined as,

\begin{equation}
    \mathcal{L}_{survival} = -\sum_{i=1}^{m} y_{i} y^{\prime}_{i} + \log \sum_{j=0}^{m} \exp\sum_{i=j+1}^{m} y^{\prime}_{i}
\end{equation}
which can be directly applied to the downstream module and included in the joint loss function of OmiEmbed.

\subsection{Multi-task strategy}
\label{sec:multitask_method}
With the joint loss function (\ref{eq:down_loss}) of the multi-task downstream modules, we aimed to train multiple downstream nets in OmiEmbed simultaneously and efficiently instead of separate training so as to obtain a unified network that can reconstruct a comprehensive phenotype profile of each subject. In order to balance the optimisation of different tasks, we adapted the multi-task optimisation method gradient normalisation (GradNorm) \citep{Chen2018GradNormGN} to our OmiEmbed framework.

In equation (\ref{eq:down_loss}) $w_k$ is the weight of each downstream loss, and the weight can also be regarded as a trainable parameter that varies at each training iteration. The idea of GradNorm is to penalise the network if gradients of any downstream task are too large or too small, which makes all the tasks train at similar rates \citep{Chen2018GradNormGN}. Firstly the gradient norm of each downstream task is calculated by 
\begin{equation}
    G_{\theta}^{(k)} = \left\|\nabla_{\theta} w_{k} \mathcal{L}_{down_k}\right\|_{2}
\end{equation}
where $\theta$ is the parameters of the last encoding layer in the deep embedding module of OmiEmbed. The average gradient norm among all tasks can be then calculated by 
\begin{equation}
    \bar{G}_{\theta} = \frac{1}{K} \sum_{k=1}^{K} G_{\theta}^{(k)} 
\end{equation}
where $K$ is the number of downstream tasks. The relative inverse training rate of each task can be defined as:
\begin{equation}
    r_{k}=\frac{\tilde{\mathcal{L}}_{down_k}} {\frac{1}{K} \sum_{k=1}^{K} \tilde{\mathcal{L}}_{down_k}}
\end{equation}
where $\tilde{\mathcal{L}}_{down_k} = \mathcal{L}_{down_k} / \mathcal{L}_{down_{k0}}$ which is the ratio of the current loss to the initial loss of the downstream task $k$. Then the loss of GradNorm can be defined as:
\begin{equation}
    \mathcal{L}_{grad}=\sum_{k=1}^{K} \left|G_{\theta}^{(k)}-\bar{G}_{\theta} \times {r_{k}} ^{\alpha}\right|_{1}
    \label{eq:gradnorm_loss}
\end{equation}
where $\alpha$ is the hyperparameter that represents strength pulling tasks back to a common training rate. A separate backpropagation process was conducted during each training iteration on $\mathcal{L}_{grad}$, which was only used for updating $w_k$.

\section{Results}

\subsection{Implementation details}
The OmiEmbed multi-omics multi-task framework was built on the deep learning library PyTorch \citep{Paszke2019PyTorchAI}. The code of OmiEmbed has been made open-source on GitHub\footnote{\url{https://github.com/zhangxiaoyu11/OmiEmbed/}}, and it is easy to apply it on any high-dimensional omics dataset for any aforementioned downstream tasks individually or collaboratively. The detailed network structures of both the FC-type and CNN-type deep embedding modules were illustrated in Supplementary Figure S1 and S2. In the FC-type omics embedding network, CpG sites of DNA methylation profiles were separately connected to different hidden layers based on their targeting chromosomes in order to reduce the number of parameters, prevent overfitting and save the GPU memory. The chromosome separation step can be automatically processed in OmiEmbed with a built-in DNA methylation annotation if the FC-type embedding was selected using corresponding command-line arguments. Other deep learning techniques were also applied to prevent overfitting in OmiEmbed including dropout \citep{srivastava2014dropout}, batch normalisation \citep{ioffe2015batch}, weight decay regularisation and the learning rate schedule. 

The model was trained on two NVIDIA Titan X GPUs with 12 gigabytes of memory each. The input dataset for each experiment was randomly separated into training, validation and testing sets. The separation was conducted in a stratified manner so as to keep the proportion of each class in all three sets. Stratified 5-fold cross-validation was also applied to robustly evaluate the performance of OmiEmbed and other compared methods avoiding bias from specific test set. The open-source project files of OmiEmbed are well-organised with clear structures and predefined classes which make it convenient to extend the framework to other customised input data, network structures and downstream tasks.

\begin{figure*}[!t]
    \centering
    \includegraphics[scale=0.45]{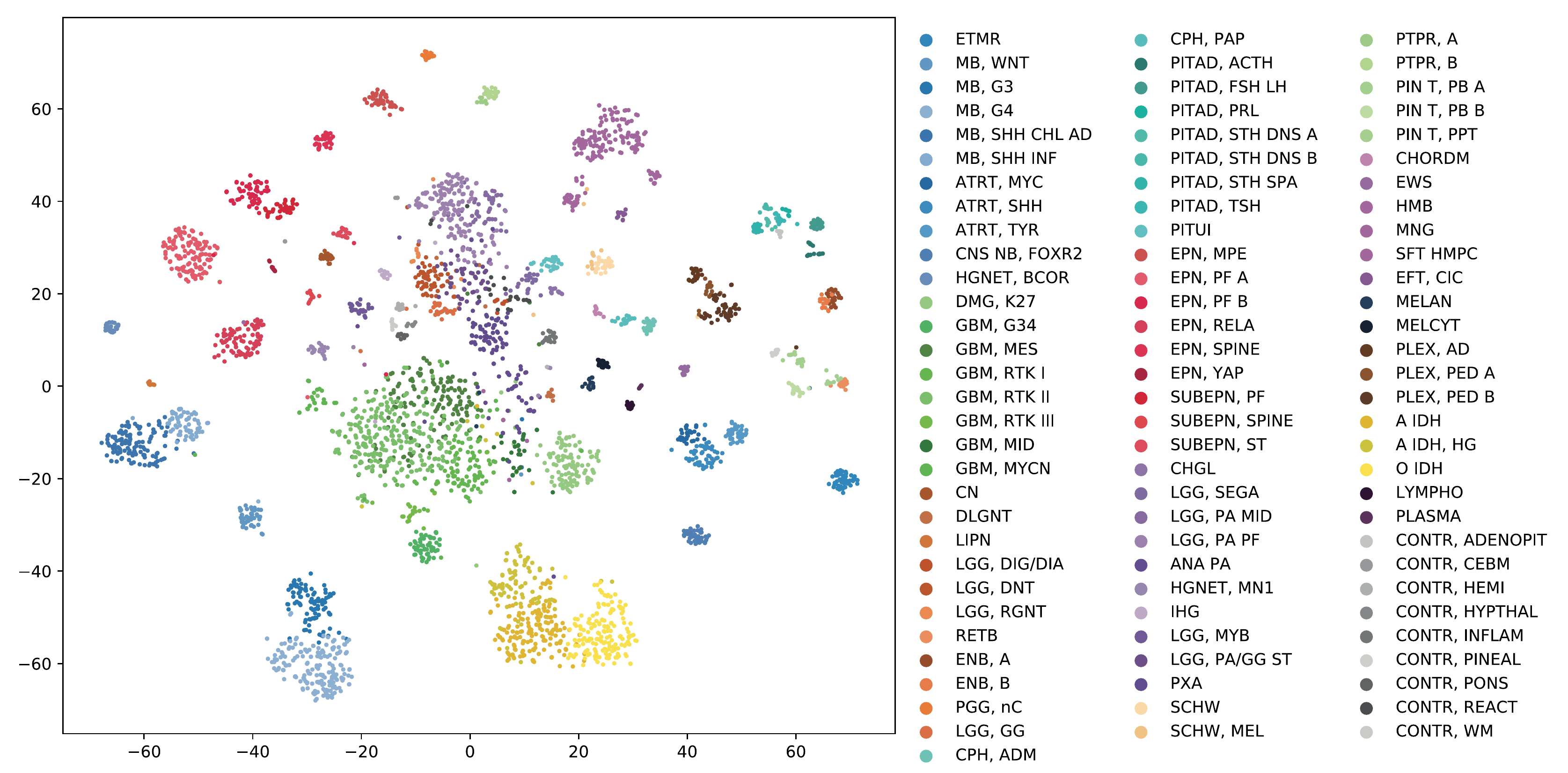}
    \caption{The 128D latent space of the BTM dataset learned by the unsupervised phase of OmiEmbed. The scatter graph was visualised using t-SNE. Each label in the scatter graph was colour by its methylation class label and the full name of each class abbreviation can be found in Supplementary Table S8. Tumour types belonging to the same upper-level class were marked in similar colours.}\label{fig:btm_scatter}
\end{figure*}

\subsection{Dimensionality reduction}
OmiEmbed can be regarded as an unsupervised dimensionality reduction method when only the training Phase 1 mentioned was applied in the experiment. The high-dimensional multi-omics data can be compressed into a new representation with the target dimensionality set by the command line argument of OmiEmbed. Then the output file can be directly used for visualisation or any other downstream tasks. Here we reduced each sample in the BTM dataset into a 128D latent vector using the unsupervised Phase 1 of OmiEmbed. The learned latent space of the BTM dataset was visualised by t-distributed stochastic neighbour embedding (t-SNE) \citep{van2008visualizing} and shown in Figure \ref{fig:btm_scatter}. Samples of different brain tumour types automatically clustered together in the latent space and tumour types belonging to the same upper-level class (e.g., glioblastoma, embryonal tumour, and ependymal tumour) also formed into the corresponding upper-level clusters.

\begin{figure}[!t]
    \includegraphics[scale=0.43]{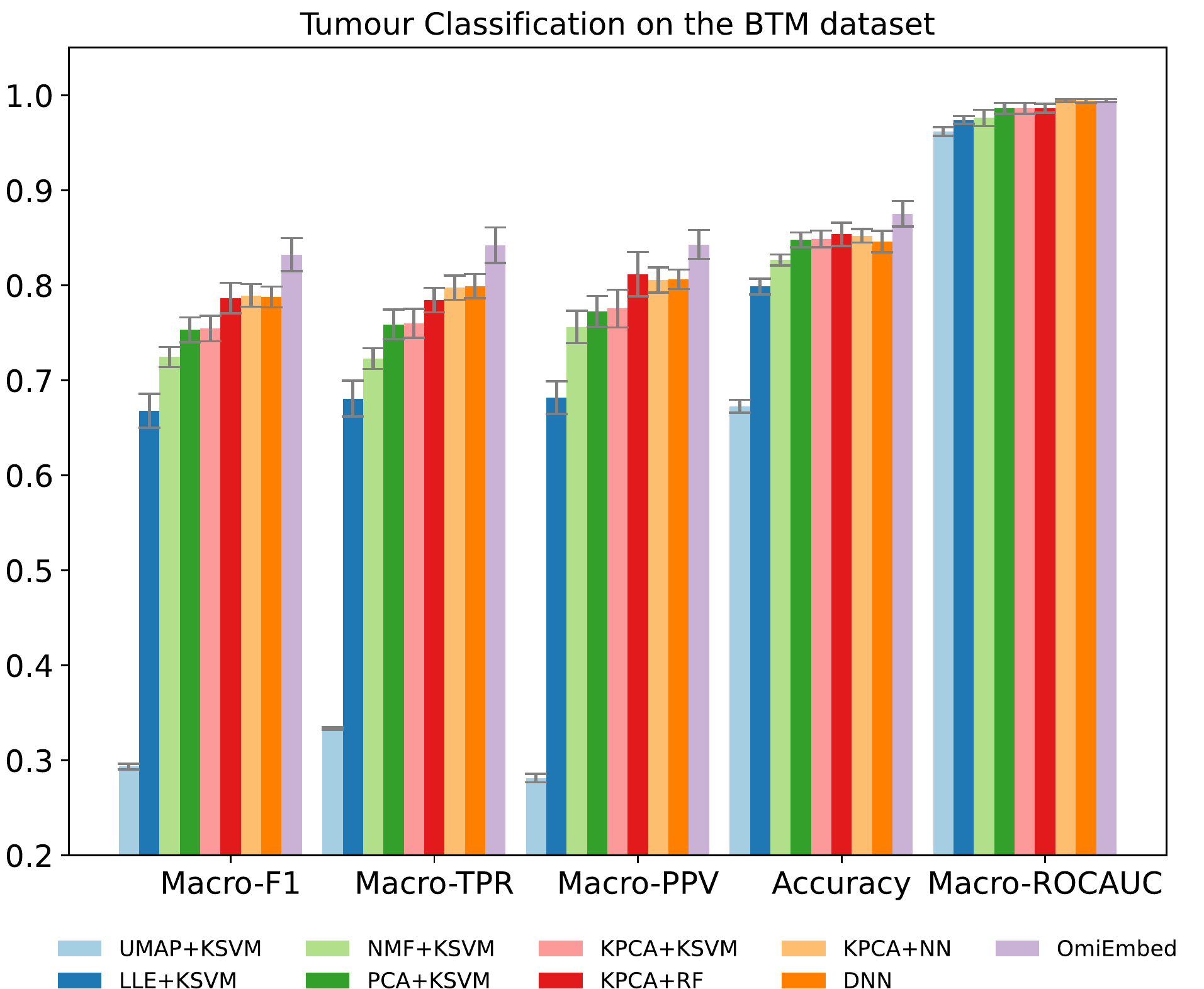}
    \centering
    \caption{Performance comparison of OmiEmbed and other eight methods for the tumour entity classification task on the BTM dataset with the histopathological tumour type labels.}
    \label{fig:pathological}
\end{figure}

\begin{table*}[t!]
    \caption{The classification performance on the BTM dataset using the histopathological tumour type labels (2016 WHO classification) with 5-fold cross-validation, which was measured by macro-averaged F1 score (Macro-F1), macro-averaged true positive rate (Macro-TPR), macro-averaged positive predictive value (Macro-PPV), overall accuracy and macro-averaged area under the receiver operating characteristic curve (Macro-ROCAUC).\label{tab:brain_who_class}}
    \tabcolsep=0pt
    \begin{tabular*}{\textwidth}{@{\extracolsep{\fill}}lcccccc@{\extracolsep{\fill}}}
        \toprule
        & Macro-F1            & Macro-TPR           & Macro-PPV           & Accuracy           & Macro-ROCAUC           \\
        \midrule
        UMAP+KSVM & 0.2933$\pm$0.0029 & 0.3334$\pm$0.0012 & 0.2812$\pm$0.0044 & 0.6725$\pm$0.0067 & 0.9619$\pm$0.0045 \\
        LLE+KSVM  & 0.6678$\pm$0.0178 & 0.6807$\pm$0.0189 & 0.6817$\pm$0.0172 & 0.7987$\pm$0.0089 & 0.9740$\pm$0.0041 \\
        NMF+KSVM  & 0.7245$\pm$0.0107 & 0.7228$\pm$0.0110 & 0.7560$\pm$0.0172 & 0.8266$\pm$0.0057 & 0.9761$\pm$0.0085 \\
        PCA+KSVM  & 0.7531$\pm$0.0131 & 0.7588$\pm$0.0155 & 0.7724$\pm$0.0163 & 0.8477$\pm$0.0078 & 0.9861$\pm$0.0058 \\
        KPCA+KSVM & 0.7544$\pm$0.0135 & 0.7598$\pm$0.0153 & 0.7755$\pm$0.0199 & 0.8488$\pm$0.0087 & 0.9863$\pm$0.0058 \\
        \midrule
        KPCA+RF   & 0.7865$\pm$0.0162 & 0.7844$\pm$0.0129 & 0.8117$\pm$0.0233 & 0.8537$\pm$0.0122 & 0.9862$\pm$0.0046 \\
        KPCA+NN   & 0.7893$\pm$0.0120 & 0.7975$\pm$0.0127 & 0.8056$\pm$0.0134 & 0.8521$\pm$0.0070 & 0.9940$\pm$0.0015 \\
        \midrule
        DNN       & 0.7877$\pm$0.0108 & 0.7991$\pm$0.0128 & 0.8063$\pm$0.0103 & 0.8459$\pm$0.0112 & 0.9940$\pm$0.0021 \\
        \textbf{OmiEmbed} & \textbf{0.8323$\pm$0.0174} & \textbf{0.8421$\pm$0.0188} & \textbf{0.8429$\pm$0.0152} & \textbf{0.8754$\pm$0.0133} & \textbf{0.9943$\pm$0.0016} \\
        \botrule
    \end{tabular*}
\end{table*}

\begin{table*}[!t]
    \caption{Three output examples from the testing set of histopathological tumour type classification on the BTM dataset.\label{tab:class_probability}}%
    \begin{tabular*}{\textwidth}{@{\extracolsep\fill}lllll@{\extracolsep\fill}}
        \toprule
        &  & Class ID & Class name (2016 WHO classification of CNS tumours) & Probability \\
        \midrule
        \multirow{6}{*}{GSM2941340} & Top 1 & 4 & Anaplastic ependymoma & 75.3695\% \\
        & Top 2 & 5 & Ependymoma & 21.5774\% \\
        & Top 3 & 23 & Myxopapillary ependymoma & 0.0275\% \\
        & Top 4 & 62 & Anaplastic (malignant) meningioma & 0.0008\% \\
        & Top 5 & 31 & Atypical meningioma & 0.0004\% \\
        & Ground Truth & 4 & Anaplastic ependymoma &  \\
        \midrule
        \multirow{6}{*}{GSM2941792} & Top 1 & 6 & Anaplastic astrocytoma, IDH-mutant & 81.4425\% \\
        & Top 2 & 13 & Diffuse astrocytoma, IDH-mutant & 17.8859\% \\
        & Top 3 & 15 & Oligodendroglioma, IDH-mutant and 1p/19q-codeleted & 0.4496\% \\
        & Top 4 & 10 & Anaplastic oligodendroglioma, IDH-mutant and 1p/19q-codeleted & 0.1553\% \\
        & Top 5 & 54 & Anaplastic astrocytoma, IDH-wildtype & 0.0124\% \\
        & Ground Truth & 6 & Anaplastic astrocytoma, IDH-mutant &  \\
        \midrule
        \multirow{6}{*}{GSM2405444} & Top 1 & 0 & Glioblastoma, IDH-wildtype & 99.9995\% \\
        & Top 2 & 32 & Gliosarcoma, IDH-wildtype & 0.0001\% \\
        & Top 3 & 44 & Anaplastic pilocytic astrocytoma & 0.0001\% \\
        & Top 4 & 52 & Anaplastic pilocytic astrocytoma (unresolved status) & 0.0001\% \\
        & Top 5 & 20 & Ganglioglioma & 0.0001\% \\
        & Ground Truth & 0 & Glioblastoma, IDH-wildtype & \\
        \botrule
    \end{tabular*}
\end{table*}

\subsection{Tumour classification}
Instead of using the training Phase 1 individually as a dimensionality reduction method and separately training the downstream task with other machine learning algorithms, using all of the three training phases of OmiEmbed in an end-to-end manner is more efficient with better results. Here we first tested the classification performance of OmiEmbed on the BTM dataset. There are two types of tumour type classification systems in this brain tumour dataset: the histopathological tumour type labels defined by the 2016 WHO classification \citep{louis20162016} and the methylation tumour type labels defined by the original paper of this dataset \citep{capper2018dna}. For each type of these two classification systems, the 3,905 samples were divided into more than 90 classes including different normal control types (e.g., normal cortex, normal pons, normal pineal gland, and normal corpus callosum).

The classification results of OmiEmbed on the BTM dataset with the histopathological labels were shown in Table \ref{tab:brain_who_class} and Figure \ref{fig:pathological}. The results with the methylation label were shown in Supplementary Table S2 and Supplementary Figure S3. The performance was measured by five multi-class classification metrics: macro-averaged F1 score (Macro-F1), macro-averaged true positive rate (Macro-TPR), macro-averaged positive predictive value (Macro-PPV), overall accuracy and macro-averaged area under the receiver operating characteristic curve (Macro-ROCAUC). TPR is also known as sensitivity or recall and PPV is also known as precision. 

The results were first compared with the combination of five dimensionality reduction methods and kernel support vector machine (KSVM). The five different dimensionality reduction methods included uniform manifold approximation and projection (UMAP) \citep{McInnes2018UMAPUM}, locally linear embedding (LLE) \citep{Roweis2000NonlinearDR}, non-negative matrix factorization (NMF), principal component analysis (PCA) and kernel principal component analysis (KPCA) \citep{scholkopf1997kernel}. The original data from the BTM dataset was first reduced to 128D by the aforementioned dimensionality reduction methods and then classified by the KSVM with a radial basis function (RBF) kernel. Best results among them achieved by KPCA, therefore, other machine learning methods were evaluated along with KPCA, including random forest (RF) \citep{Breiman2004RandomF} and neural network (NN). The NN used here was comprised of two hidden layers with 128 neurons and 64 neurons respectively. The deep neural network (DNN) with the structure of 1024-512-256-128 was also compared with OmiEmbed in an end-to-end manner. As illustrated in Table \ref{tab:brain_who_class}, Figure \ref{fig:pathological}, Supplementary Table S2 and Supplementary Figure S3, OmiEmbed achieved the best classification performance in all the five metrics with both types of classification systems. 

In stratified medicine, the probability of certain diagnostic prediction is as important as the prediction itself. With the softmax layer in the classification downstream task module, OmiEmbed is able to output the predicted diagnosis as well as the probability of every class for each input sample. Table \ref{tab:class_probability} demonstrated three output examples from the testing set of the histopathological tumour type classification task on the BTM dataset. For testing samples like GSM2941340 and GSM2941792 which are difficult to determine the tumour type, OmiEmbed is able to not only predict the correct diagnosis but detect analogous tumour entities (e.g., Anaplastic astrocytoma IDH-mutant and Diffuse astrocytoma IDH-mutant) and rank them by the class probability.

\begin{table*}[t!]
    \caption{The performance of tumour type classification on the GDC multi-omics dataset with different omics type combinations.\label{tab:pancan_tumour_type_multiomics}}
    \tabcolsep=0pt
    \begin{tabular*}{\textwidth}{@{\extracolsep{\fill}}lccccc@{\extracolsep{\fill}}}
        \toprule 
        & Macro-F1 & Macro-TPR & Macro-PPV & Accuracy & Macro-ROCAUC \\
        \midrule
        Gene expression (a) & 0.9518$\pm$0.0053 & 0.9522$\pm$0.0055 & 0.9558$\pm$0.0069 & 0.9676$\pm$0.0027 & 0.9982$\pm$0.0003 \\
        DNA methylation (b) & 0.9273$\pm$0.0167 & 0.9253$\pm$0.0192 & 0.9333$\pm$0.0181 & 0.9650$\pm$0.0040 & 0.9985$\pm$0.0001 \\
        miRNA expression (c) & 0.9274$\pm$0.0140 & 0.9268$\pm$0.0134 & 0.9320$\pm$0.0148 & 0.9544$\pm$0.0057 & 0.9983$\pm$0.0004 \\
        Multi-omics (a+b) & 0.9675$\pm$0.0083 & 0.9669$\pm$0.0077 & 0.9643$\pm$0.0077 & 0.9753$\pm$0.0040 & 0.9988$\pm$0.0009 \\
        \textbf{Multi-omics (a+b+c)} & \textbf{0.9683$\pm$0.0020} & \textbf{0.9684$\pm$0.0026} & \textbf{0.9705$\pm$0.0047} & \textbf{0.9771$\pm$0.0027} & \textbf{0.9991$\pm$0.0002} \\
        \botrule
    \end{tabular*}
\end{table*}

\subsection{Multi-omics integration}
Different types of omics profiles can be integrated into single latent representation and used for different downstream tasks through the multi-omics deep embedding module of OmiEmbed. In order to test the effect of multi-omics integration on the downstream task, Tumour type classifiers were trained on the GDC multi-omics dataset using OmiEmbed. Three types of omics data in the GDC dataset were used in the experiments: RNA-Seq gene expression, DNA methylation and miRNA expression. There are 33 tumour types and normal control class (34 classes in total) in the dataset. We trained the model with each omics type alone and two different multiple omics type combinations. The classification performance in each scenario was shown in Table \ref{tab:pancan_tumour_type_multiomics}. The performance metrics for each omics type alone were close to each other and the best metrics were achieved with the combination of all three omics types. This result also indicates combining multiple omics data can yield better insights into the underlying mechanisms of diseases.

\begin{table*}[t!]
    \caption{The classification performance of predicting four categorical phenotype features on the GDC dataset.\label{tab:pancan_other_pheno}}
    \tabcolsep=0pt
    \begin{tabular*}{\textwidth}{@{\extracolsep{\fill}}lccccccc@{\extracolsep{\fill}}}
        \toprule 
        &     & Macro-F1      & Macro-TPR     & Macro-PPV     & Accuracy      & Macro-ROCAUC  \\
        \midrule
        \multirow{2}{*}{Disease stage} 
         & DNN & 0.7530$\pm$0.0035 & 0.7552$\pm$0.0090 & 0.7517$\pm$0.0105 & 0.9782$\pm$0.0014 & 0.9552$\pm$0.0207 \\
         & \textbf{OmiEmbed} & \textbf{0.8173$\pm$0.0401} & \textbf{0.8016$\pm$0.0291} & \textbf{0.8610$\pm$0.0816} & \textbf{0.9797$\pm$0.0024} & \textbf{0.9540$\pm$0.0320} \\
        \midrule
        \multirow{2}{*}{Primary site} 
         & DNN & 0.9639$\pm$0.0066 & 0.9638$\pm$0.011 & 0.9576$\pm$0.0123 & 0.9593$\pm$0.0116 & 0.9987$\pm$0.0006 \\
         & \textbf{OmiEmbed} & \textbf{0.9717$\pm$0.0066} & \textbf{0.9711$\pm$0.0046} & \textbf{0.9734$\pm$0.0095} & \textbf{0.9812$\pm$0.0023} & \textbf{0.9994$\pm$0.0003} \\
        \midrule
        \multirow{2}{*}{Race} 
         & DNN & 0.4702$\pm$0.1455 & 0.4644$\pm$0.1337 & 0.4799$\pm$0.1580 & 0.9353$\pm$0.0567 & 0.7715$\pm$0.0765 \\
         & \textbf{OmiEmbed} & \textbf{0.6124$\pm$0.0946} & \textbf{0.5960$\pm$0.0764} & \textbf{0.6394$\pm$0.1349} & \textbf{0.9767$\pm$0.0050} & \textbf{0.9068$\pm$0.0266} \\
        \midrule
        \multirow{2}{*}{Gender} 
         & DNN & 0.8701$\pm$0.0725 & 0.887$\pm$0.0413 & 0.8713$\pm$0.0705 & 0.8668$\pm$0.079 & 0.962$\pm$0.0102 \\
         & \textbf{OmiEmbed} & \textbf{0.9560$\pm$0.0023} & \textbf{0.9558$\pm$0.0024} & \textbf{0.9568$\pm$0.0018} & \textbf{0.9561$\pm$0.0022} & \textbf{0.9903$\pm$0.0019} \\
        \botrule
    \end{tabular*}
\end{table*}

\begin{figure}[!t]
    \includegraphics[scale=0.4136]{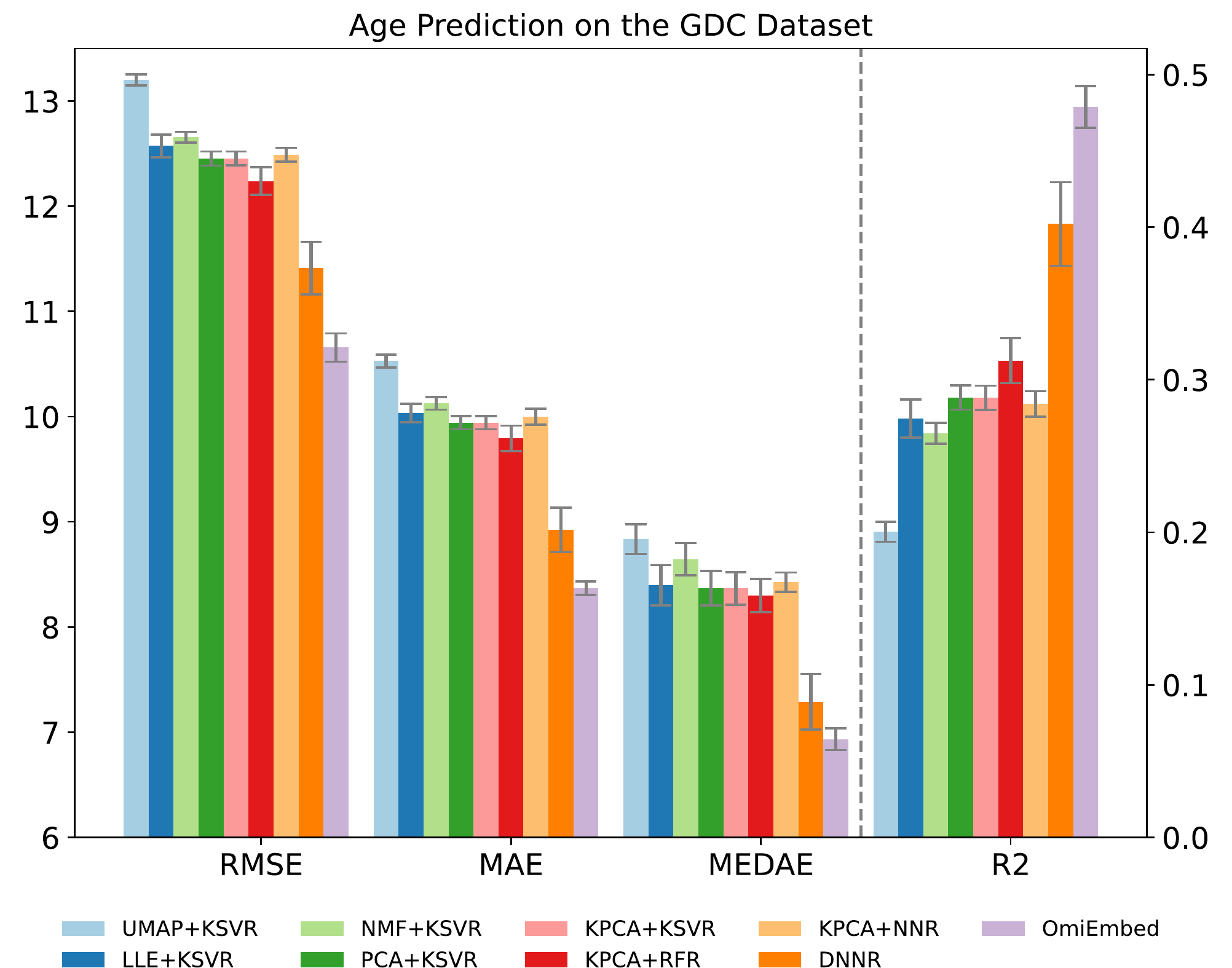}
    \centering
    \caption{Performance comparison of OmiEmbed and other eight methods for the age prediction task on the GDC dataset. For root mean square error (RMSE), mean absolute error (MAE) and median absolute error (MEDAE), lower values mean better regression performance. For $\rm R^2$ score, higher values mean better regression performance.}
    \label{fig:age}
\end{figure}

\begin{table*}[t!]
    \caption{The age prediction performance of OmiEmbed and eight other methods on the GDC dataset. For median absolute error, mean absolute error and RMSE, lower values mean better regression performance. For $\rm R^2$ score, higher values mean better regression performance. \label{tab:pancan_age}}
    \tabcolsep=0pt
    \begin{tabular*}{\textwidth}{@{\extracolsep{\fill}}lccccc@{\extracolsep{\fill}}}
        \toprule
        & Median Absolute Error & Mean Absolute Error & RMSE           & $\rm R^2$            \\
        \midrule
        UMAP+KSVR & 8.8353$\pm$0.1408         & 10.5292$\pm$0.0621      & 13.2020$\pm$0.0519 & 0.2003$\pm$0.0065 \\
        LLE+KSVR  & 8.3977$\pm$0.1910         & 10.0348$\pm$0.0885      & 12.5742$\pm$0.1075 & 0.2745$\pm$0.0125 \\
        NMF+KSVR  & 8.6448$\pm$0.1534         & 10.1268$\pm$0.0589      & 12.6571$\pm$0.0498 & 0.2649$\pm$0.0069 \\
        PCA+KSVR  & 8.3691$\pm$0.1646         & 9.9430$\pm$0.0633       & 12.4547$\pm$0.0669 & 0.2882$\pm$0.0080 \\
        KPCA+KSVR & 8.3674$\pm$0.1544         & 9.9432$\pm$0.0633       & 12.4561$\pm$0.0666 & 0.2881$\pm$0.0080 \\
        \midrule
        KPCA+RFR  & 8.2990$\pm$0.1576         & 9.7939$\pm$0.1225       & 12.2403$\pm$0.1307 & 0.3125$\pm$0.0148 \\
        KPCA+NNR  & 8.4269$\pm$0.0917         & 9.9995$\pm$0.0752       & 12.4907$\pm$0.0649 & 0.2841$\pm$0.0083 \\
        \midrule
        DNNR      & 7.2897$\pm$0.2649         & 8.9247$\pm$0.2101       & 11.4128$\pm$0.2490 & 0.4020$\pm$0.0273 \\
        \textbf{OmiEmbed} & \textbf{6.9330$\pm$0.1031} & \textbf{8.3694$\pm$0.0648} & \textbf{10.6578$\pm$0.1337} & \textbf{0.4788$\pm$0.0137}\\
        \botrule
    \end{tabular*}
\end{table*}

\subsection{Reconstruction of demographic and clinical features}
With both the classification and regression downstream networks built in OmiEmbed, we were able to reconstruct a number of phenotype features from high-dimensional omics data. Here we tested the prediction performance of four different phenotype features in the GDC dataset including age, gender, race, the disease stage and primary site of the clinical sample. Detailed information of each categorical features was listed in Supplementary Table S3. 

The disease stage is the clinical type of the sample, which consists of primary tumour, metastatic tumour, recurrent tumour and normal control tissue. The primary site is the place where the cancer starts growing. Samples in the GDC dataset are from 28 different primary sites such as breast, kidney, lung, and skin. The race of the subjects in the GDC dataset was divided into six main categories: white, Asian, black or African American, American Indian or Alaska native, native Hawaiian or other Pacific islander and others. As for the gender of each sample, since the molecular features targeting the Y chromosome were filtered in the preprocessing stage, the model was required to classify the gender based on other molecular features. The OmiEmbed classification performance of the four categorical phenotype features was shown in Table \ref{tab:pancan_other_pheno} along with the results of a DNN with the structure of 1024-512-256-128. The full comparison of all nine methods was illustrated in Supplementary Table S4-S7 and Supplementary Figure S4-S7.  

Since the label of age is numerical instead of categorical, the regression downstream module was applied for the age prediction task. The performance of age prediction was evaluated by the three regression metrics: median absolute error, mean absolute error, root mean square error (RMSE) and coefficient of determination ($\rm R^2$) which was illustrated in Table \ref{tab:pancan_age} and Figure \ref{fig:age}. For median absolute error, mean absolute error and RMSE lower values represent better regression performance, whereas for $\rm R^2$ score higher values indicate better regression performance.

The age prediction performance of OmiEmbed was first compared with the combination of the five aforementioned dimensionality reduction methods and kernel support vector regressor (KSVR). The original data from the GDC dataset was first reduced to 128D and then fed to the KSVR with the RBF kernel. Other regression methods were also evaluated along with KPCA, including random forest regressor (RFR) \citep{Breiman2004RandomF} and neural network regressor (NNR). The NNR adopted here was comprised of two hidden layers with 128 neurons and 64 neurons respectively. The deep neural network regressor (DNNR) with the structure of 1024-512-256-128 was also compared with OmiEmbed in an end-to-end manner. OmiEmbed achieved the best regression performance with the lowest distance error and highest coefficient of determination.

\begin{figure}[!t]
    \includegraphics[scale=0.39]{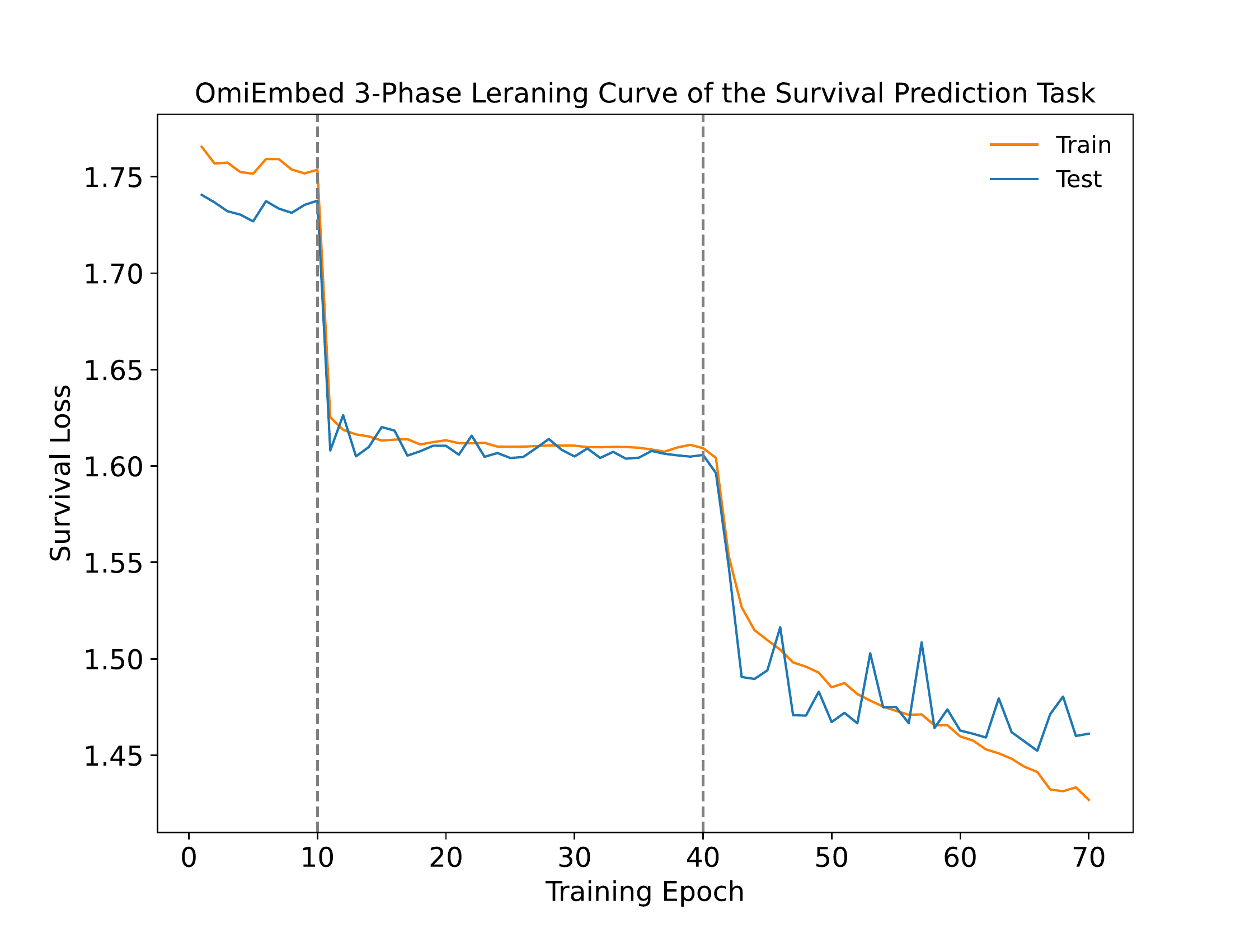}
    \centering
    \caption{The learning curve of the survival prediction task with the three-phase training strategy of OmiEmbed. Epoch 1 to 10 belong to Phase 1; epoch 11 to 40 belong to Phase 2; epoch 41 to 70 belong to Phase 3.}
    \label{fig:loss}
\end{figure}

\begin{table}[t!]
    \caption{The survival prediction performance of OmiEmbed and nine other methods on the GDC dataset. For C-index, higher values mean better prediction performance. For IBS score, lower values mean better prediction performance. \label{tab:pancan_survival}}
    \tabcolsep=0pt
    \begin{tabular*}{\columnwidth}{@{\extracolsep{\fill}}lccc@{\extracolsep{\fill}}}
        \toprule
        & C-index                & IBS                    \\
        \midrule
        UMAP+CoxPH          & 0.6986$\pm$0.0179          & 0.1847$\pm$0.0258          \\
        LLE+CoxPH           & 0.6833$\pm$0.0215          & 0.2108$\pm$0.0255          \\
        NMF+CoxPH           & 0.7070$\pm$0.0215          & 0.1822$\pm$0.0277          \\
        PCA+CoxPH           & 0.7097$\pm$0.0228          & 0.1783$\pm$0.0248          \\
        KPCA+CoxPH          & 0.7096$\pm$0.0233          & 0.1778$\pm$0.0244          \\
        \midrule
        KPCA+RSF            & 0.6854$\pm$0.0244          & 0.1896$\pm$0.0104          \\
        KPCA+CSF            & 0.6909$\pm$0.0255          & 0.1906$\pm$0.0116          \\
        KPCA+EST            & 0.6931$\pm$0.0286          & 0.1938$\pm$0.0115          \\
        \midrule
        DeepSurv            & 0.7180$\pm$0.0171          & 0.2468$\pm$0.0269          \\
        \textbf{OmiEmbed}   & \textbf{0.7715$\pm$0.0073} & \textbf{0.1657$\pm$0.0224} \\
        \botrule
    \end{tabular*}
\end{table}

\begin{figure}[!t]
    \includegraphics[scale=0.39]{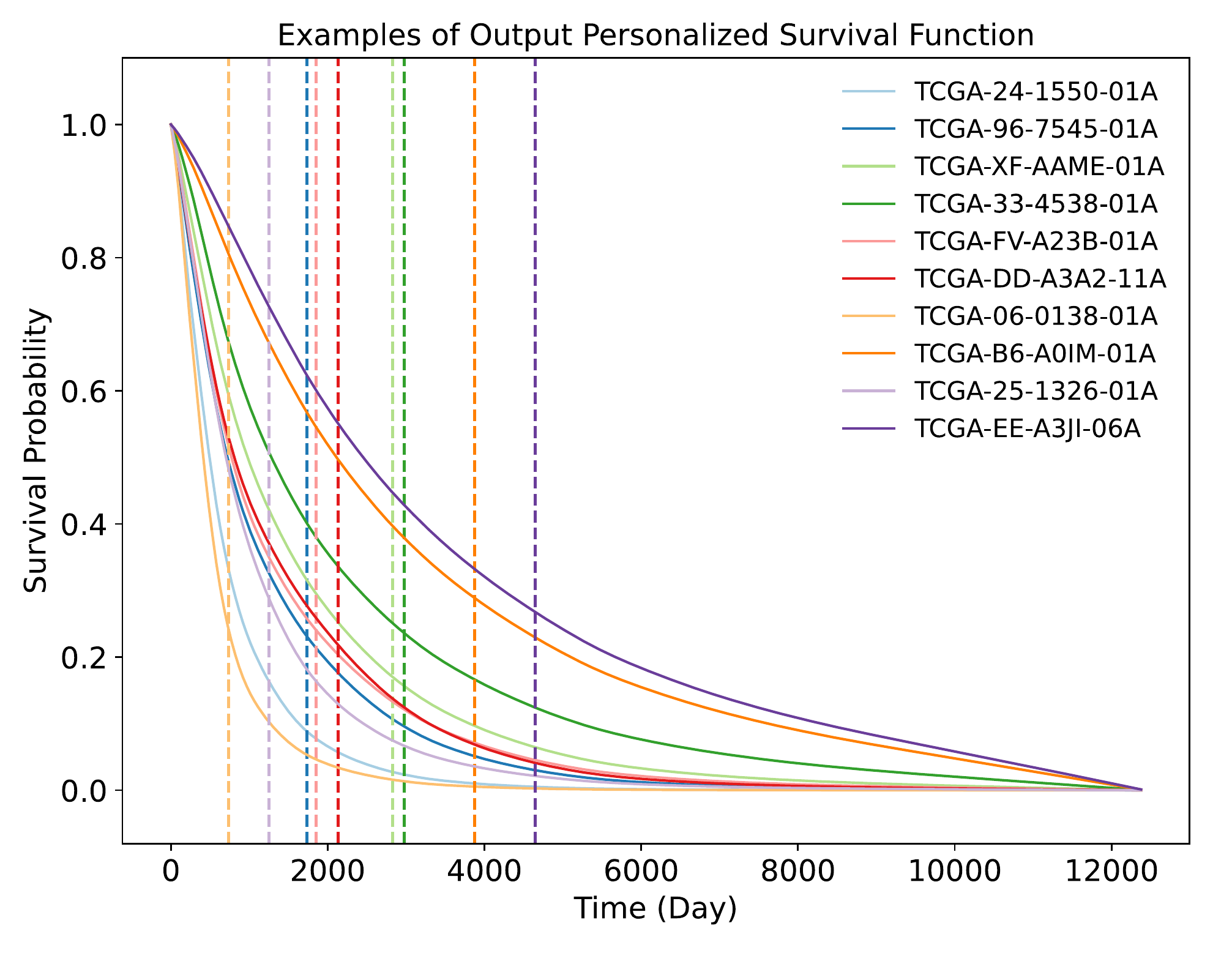}
    \centering
    \caption{Personalised survival curves of ten random subjects from the testing set of the GDC dataset. The dashed vertical line with the corresponding colour indicates the death time of each subject.}
    \label{fig:survival_curve}
\end{figure}

\begin{figure}[!t]
    \includegraphics[scale=0.40]{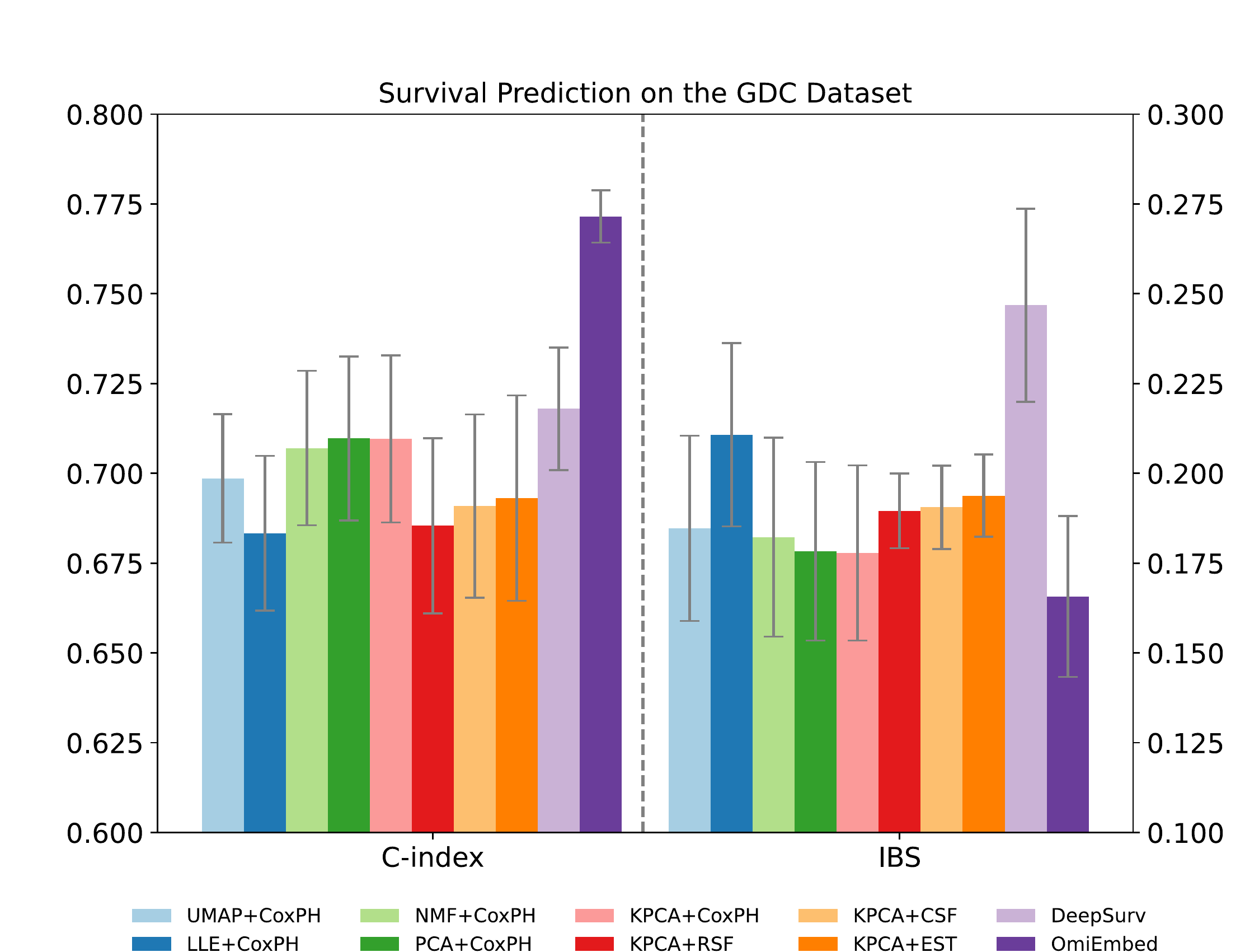}
    \centering
    \caption{Performance comparison of OmiEmbed and other nine methods for the survival prediction task on the GDC dataset. For C-index, higher values mean better prediction performance. For IBS score, lower values mean better prediction performance.}
    \label{fig:survival}
\end{figure}

\subsection{Survival prediction}
With the survival prediction downstream module of OmiEmbed, we are able to predict the survival function of each subject from corresponding high-dimensional omics data. Just like other downstream tasks, OmiEmbed was trained by the three-phase training strategy for the survival prediction task. Survival losses at each epoch on the training and testing set were illustrated in Figure \ref{fig:loss} with a step shape learning curve. The first ten epochs were in Phase 1 where the embedding network was pretrained in an unsupervised manner. In epoch 11 to epoch 40, the downstream network was trained individually when the pre-trained embedding network was fixed, and in the last phase the whole network was fine-tuned for better performance, which is consistent with the learning curve.

The performance of the survival prediction downstream task was evaluated by concordance index (C-index) and integrated Brier score (IBS) which are the most commonly used metrics for survival prediction. A C-index value of 1 indicates the perfect prediction model and a value of 0.5 signifies that the performance of the model is similar to expected at random. The Brier score indicates the accuracy of a predicted survival function at a certain time point, which is between 0 and 1. IBS is the average Brier score among all available times providing an overall calculation of the model performance. 

The results of OmiEmbed was compared with methods that first reduced the dimensionality of input omics data to 128D using UMAP \citep{McInnes2018UMAPUM}, LLE \citep{Roweis2000NonlinearDR}, NMF, PCA or KPCA \citep{scholkopf1997kernel} and then fed the 128D latent vectors to the survival prediction method Cox Proportional Hazard model (CoxPH). Other survival prediction methods, including random survival forest (RSF) \citep{Ishwaran2008RandomSF}, conditional survival forest (CSF) \citep{Wright2017UnbiasedSV} and extra survival trees (EST) \citep{Geurts2006ExtremelyRT}, were also evaluated after reduced to 128D latent vectors by KPCA. The survival prediction performance of OmiEmbed was also compared with the state-of-the-art deep learning method DeepSurv \citep{Katzman2018DeepSurvPT}. OmiEmbed got the best C-index (0.7715) and IBS (0.1657) among the other nine methods as shown in Table \ref{tab:pancan_survival} and Figure \ref{fig:survival}.

OmiEmbed is able to output the personalised survival function based on the corresponding omics profile. As illustrated in Figure \ref{fig:survival_curve} we randomly selected ten subjects with their observed death time from the testing set of the GDC dataset as examples and plotted the survival curve for each of them. The actual death time of each subject was also marked in the figure by the dashed vertical line with the corresponding colour.

\begin{table*}[t!]
    \caption{Multi-task training performance of OmiEmbed with three typical downstream tasks. For C-index, macro-F1 and accuracy, higher values mean better prediction performance. For IBS score, RMSE and median absolute error, lower values mean better prediction performance.\label{tab:pancan_multitask}}
    \tabcolsep=0pt
    \begin{tabular*}{\textwidth}{@{\extracolsep{\fill}}lcccccc@{\extracolsep{\fill}}}
        \toprule
        & \multicolumn{2}{c}{Survival} & \multicolumn{2}{c}{Tumour Type} & \multicolumn{2}{c}{Age}      \\
        \cline{2-3}\cline{4-5}\cline{6-7}
        & C-index         & IBS        & Macro-F1        & Accuracy        & RMSE & Median Absolute Error \\      
        \midrule
        Single task alone & 0.7715$\pm$0.0073 & 0.1657$\pm$0.0224 & 0.9518$\pm$0.0053 & 0.9676$\pm$0.0027 & 10.6578$\pm$0.1337 & 6.9330$\pm$0.1031 \\
        \textbf{Multi-task} & \textbf{0.7823$\pm$0.0076} & \textbf{0.1590$\pm$0.0212} & \textbf{0.9653$\pm$0.0057} & \textbf{0.9733$\pm$0.0029} & \textbf{10.6336$\pm$0.1034} & \textbf{6.6759$\pm$0.1195} \\
        \botrule
    \end{tabular*}
\end{table*}

\begin{figure*}[!t]
    \includegraphics[scale=0.46]{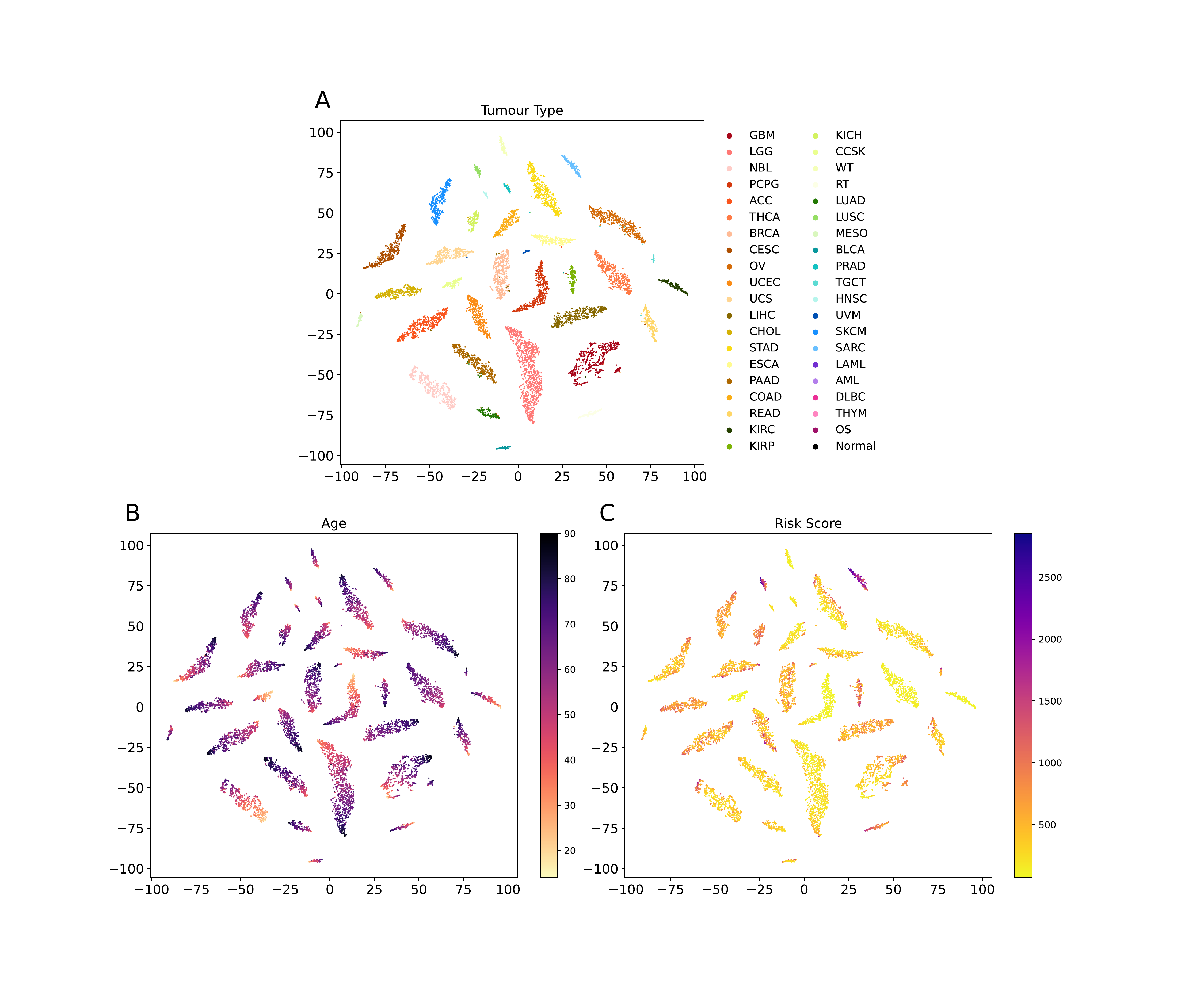}
    \centering
    \caption{Visualisation of the latent space learned by the multi-task OmiEmbed. Each sample was colour by its tumour type (A), age (B) and risk score (C). Apparent patterns related to the labels can be seen in the three scatter graphs.}
    \label{fig:latent}
\end{figure*}

\subsection{Multi-task learning}
Instead of training each aforementioned downstream module separately, we can also train multiple downstream modules together using the multi-task training strategy expatiated in the previous section. With the multi-task strategy, OmiEmbed is able to perform diverse downstream tasks simultaneously and reconstruct a comprehensive phenotype profile of each subject from high-dimensional omics data using one unified network in one forward propagation. In order to test the multi-task performance of OmiEmbed, we first selected three typical downstream tasks belonging to three distinct categories, the survival prediction task, the tumour type classification task and the age regression task, for the evaluation. Three downstream modules along with the deep embedding module were trained together using the joint loss function equation (\ref{eq:total_loss}) and the GradNorm loss function equation (\ref{eq:gradnorm_loss}). As shown in Table \ref{tab:pancan_multitask}, the performance is higher in all three downstream tasks when they were trained in a unified multi-task OmiEmbed network compared with being trained separately.

Since different downstream modules shared the common deep embedding module in the multi-task learning strategy, the latent space learnt by multi-task OmiEmbed contained information from each downstream task. The learnt embedding reduce the dimensionality of each sample in the GDC dataset, which was than then visualised t-SNE and illustrated in Figure \ref{fig:latent}. Each sample was coloured by its tumour type, age and risk score in three scatter graphs, which shown apparent patterns related to the labels.

\section{Conclusion}
OmiEmbed is an open-source deep learning framework designed for multi-omics data analysis with tasks including dimensionality reduction, multi-omics data integration, tumour type classification, disease stage prediction, demographic label reconstruction and prognosis prediction. All of the aforementioned tasks can be performed collaboratively or individually by a unified architecture which is comprised of the deep embedding and downstream task modules. OmiEmbed achieved promising results in each downstream task outperforming state-of-the-art methods, and got better performance with the multi-task strategy comparing to training them individually. Our results also indicate that OmiEmbed is able to reconstruct a comprehensive profile of each subject including demographic, diagnostic and prognostic information from the multi-omics data, which has a great potential to facilitate more accurate and personalised clinical decision making. OmiEmbed is publicly available with clear structures and predefined classes, which makes the unified framework applicable to any omics type and downstream task with minimal modification. We believe OmiEmbed will also become a framework for other researchers to analyse high-dimensional omics data using the deep learning and multi-task learning methodology.


\section{Key Points}
\begin{itemize}
    \item OmiEmbed is a unified multi-task deep learning framework for omics data, supporting dimensionality reduction, multi-omics integration, tumour type classification, phenotypic feature reconstruction and survival prediction.
    \item OmiEmbed outperformed state-of-the-art methods on all three types of downstream tasks: classification, regression and survival prediction. 
    \item OmiEmbed achieved better performance using the multi-task training strategy comparing to training each downstream task individually.
    \item OmiEmbed learnt comprehensive omics embedding containing information from multiple tasks.
    \item OmiEmbed is open-source, well-organised and convenient to extend to other customised input data, network structures and downstream tasks.
\end{itemize}

\section{Availability}
The source code, user guide document, and built-in annotation file have been made publicly available on GitHub\footnote{\url{https://github.com/zhangxiaoyu11/OmiEmbed/}}. The BTM dataset is available from GEO\footnote{\url{https://www.ncbi.nlm.nih.gov/geo/query/acc.cgi?acc=GSE109381}} with the accession ID GSE109381. The harmonised GDC pan-cancer dataset can be downloaded from the UCSC Xena data portal\footnote{\url{https://xenabrowser.net/datapages/}}.

\section{Supplementary data}
Supplementary is available at GitHub\footnote{\url{https://github.com/zhangxiaoyu11/OmiEmbed/blob/main/documents/supplementary.pdf}}.

\section{Author contributions statement}
X.Z. conceived and designed the study; X.Z. prepared and preprocessed the data; X.Z. developed the framework and performed the computational analysis; X.Z. wrote the manuscript and Y.X., K.S., Y.G. helped revise it; Y.X. helped plot the tables and figures; Y.G. supervised the project. All authors read and approved the final version of the manuscript.

\section{Acknowledgments}
This work was supported by the European Union's Horizon 2020 research and innovation programme under the Marie Sk\l{}odowska-Curie grant agreement [764281].

\section{Competing interests}
The authors declare that they have no conflict of interest.

\bibliographystyle{abbrvnat}
\bibliography{reference}


\begin{biography}{}{\author{Xiaoyu Zhang} is currently a PhD candidate at Data Science Institute, Imperial College London, London, UK.}
\end{biography}

\begin{biography}{}{\author{Yuting Xing} is currently a PhD candidate at Data Science Institute, Imperial College London, London, UK.}
\end{biography}

\begin{biography}{}{\author{Kai Sun} is currently the acting operations manager of Data Science Institute, Imperial College London, London, UK.}
\end{biography}

\begin{biography}{}{\author{Yike Guo} is currently the co-director of Data Science Institute, Imperial College London, London, UK and vice-president of Hong Kong Baptist University, Hong Kong, China.}
\end{biography}

\end{document}